\documentclass[twocolumn,showpacs,amsmath,amssymb]{revtex4}

\usepackage{graphicx}
\usepackage{dcolumn}
\usepackage{bm}

\begin{document}

\title{$\alpha$-nucleus potential for $\alpha$-decay and
sub-barrier fusion}

\author{V. Yu. Denisov$^{1,2}$ and H. Ikezoe$^{1}$ }
\address{%
$^{1}$ Department of Materials Science, Japan Atomic Energy
Research Institute, Tokai, Ibaraki 319-1195, Japan \\ $^{2}$
Institute for Nuclear Research, Prospect Nauki 47,
03680 Kiev, Ukraine }%

\date{\today}

\begin{abstract}
The set of parameters for $\alpha$-nucleus potential is derived by
using the data for both the $\alpha$-decay half-lives and the
fusion cross-sections around the barrier for reactions
$\alpha$+$^{40}$Ca, $\alpha$+$^{59}$Co, $\alpha$+$^{208}$Pb. The
$\alpha$-decay half-lives are obtained in the framework of a
cluster model using the WKB approximation. The evaluated
$\alpha$-decay half-lives and the fusion cross-sections agreed
well with the data. Fusion reactions between $\alpha$-particle and
heavy nuclei can be used for both the formation of very heavy
nuclei and spectroscopic studies of the formed compound nuclei.
\end{abstract}


\maketitle

\section{Introduction}

Knowledge of the $\alpha$-nucleus interaction potential is a key
for the analysis of various reactions between $\alpha$-particle
and nuclei. By using the potential between $\alpha$-particle and
nuclei we can evaluate the cross sections for various reactions.

The $\alpha$-decay process involves sub-barrier penetration of
$\alpha$-particles through the barrier, caused by interaction
between $\alpha$-particle and nucleus. Therefore, $\alpha$-decay
half-lives depend strongly on the $\alpha$-nucleus interaction
potential too.

The fusion reaction between $\alpha$-particle and nucleus proceeds
in the opposite direction of the $\alpha$-decay reaction. However,
the same $\alpha$-nucleus interaction potential is the principal
factor to describe both reactions. Therefore it is natural to use
data for both the $\alpha$-decay half-lives and the sub-barrier
fusion reactions for determination of the $\alpha$-nucleus
interaction potential. Note that a combination of these data has
not yet been used for evaluation of $\alpha$-nucleus potential.

The nucleus-nucleus interaction potential consists of both Coulomb
repulsion and nuclear attraction parts. These two parts form a
barrier at small distances between $\alpha$-particle and nuclei.
The Coulomb component of the potential is well-known. In contrast,
the nuclear part of the potential is less well-defined. There are
many different approaches to the nuclear part of the interaction
potential between $\alpha$-particle and nuclei
\cite{huizenga,pot-nolte,back,atzrott,blendowske,mohr,royer,pot-demetriou,pot-avrigeanu,basu,ripl}.
$\alpha$-decay \cite{back,blendowske,mohr,royer,basu} and various
scattering
\cite{huizenga,pot-nolte,atzrott,pot-demetriou,pot-avrigeanu} data
are used for evaluation of the $\alpha$-nucleus potential.
However, there are no global potentials between $\alpha$-particle
and nuclei that fit with good accuracy various reaction data from
many nuclei at collision energies deeply below and around the
barrier (e.g., the IAEA Reference Input Parameter Library
\cite{ripl}). Potentials
\cite{huizenga,pot-nolte,back,atzrott,blendowske,mohr,royer,pot-demetriou,pot-avrigeanu,basu,ripl}
evaluated for the same colliding system using different approaches
differ considerably. Thus, there is a need to reduce the
uncertainty of the interaction potential around the point where
the $\alpha$-particle and nucleus touch at low collision energies.

The fusion reaction between the $\alpha$-particle and nucleus at
collision energies around the barrier is very sensitive to the
behavior of the potential around the barrier. The energy released
in $\alpha$-decay transition from the ground-state to the
ground-state is the $Q$-value. The $Q$-value of $\alpha$-decay is
smaller than the energies used in sub-barrier fusion reactions.
Therefore, the ground-state to ground-state $\alpha$-decay of
nuclei is sensitive to the values of potential over a very wide
range of distances, from close to touching to very large. The
distance between the $\alpha$-particle and nucleus is reduced
during the fusion process, while the distance increases in the
case of $\alpha$-decay of a nucleus. Thus, these reactions are the
inverse of each other, and we should describe both reactions using
the same potential. Therefore data sets for both fusion and
$\alpha$-decay together present a unique opportunity for accurate
determination of the $\alpha$-nucleus potential at the energy
range from close to zero to around the barrier. The knowledge of
the $\alpha$-nucleus potential over this energy range is also very
important for various other applications. For example, the
evaluation of the $\alpha$-particle capture rate is very important
for description of reactions in the stars
\cite{mohr,pot-demetriou,star}.

The low-energy fusion reactions and $\alpha$-decays are only
related to the real part of the potential. However, cross-sections
of various reactions at collision energies higher than the barrier
depend on the both real and imaginary parts of the
$\alpha$-nucleus potential
\cite{pot-nolte,atzrott,mohr,pot-demetriou,pot-avrigeanu}. So, the
fusion and $\alpha$-decay reactions present a unique opportunity
to reduce the number of fitting parameters using in the
determination of the real part of the $\alpha$-nucleus potential.

The experimental information on $\alpha$-decay half-lives is
extensive and is being continually updated (see Refs.
\cite{back,royer,brazil,danevich,jaeri,she,armbruster,she-xu} and
papers cited therein). The theory of $\alpha$-decay was formulated
by Gamow a long time ago \cite{gamov}. Subsequently various
microscopic
\cite{micr,kadmenski-furman,kadmeski,stewart,delion1,delion2,bm},
macroscopic cluster \cite{back,blendowske,royer,basu,she-xu} and
fission \cite{brazil,poenaru} approaches to the description of
$\alpha$-decay have been proposed. The simple empirical relations
described the $\alpha$-decay half-lives
\cite{royer,poenaru,sss,brown} are discussed also. Below we will
use a cluster approach to the $\alpha$-decay, which is the most
suitable for determination of the interaction potential between
$\alpha$-particle and nucleus. Using this potential we
simultaneously describe the available data for both the
alpha-decay half-lives and the sub-barrier fusion reaction
cross-sections.

Many $\alpha$-emitters are deformed. Therefore $\alpha$-nucleus
potential should depend on the angle $\theta$ between the
direction of $\alpha$-emission and the axial-symmetry axis of the
deformed nucleus. Both the $\alpha$-decay width and the
transmission coefficient for tunnelling through the barrier are
strongly dependent on $\theta$
\cite{micr,kadmeski,stewart,delion1,delion2,bm,aleshin,alexander,huizenga2,severijns}.
This effect is considered in detail in microscopic models
\cite{stewart,delion1,delion2,bm}. Unfortunately, deformation and
angle effects have not been considered in previously discussed
cluster models of $\alpha$-decay half-lives
\cite{back,blendowske,mohr,royer,basu,brazil,she-xu}. Below, we
take into account the deformation of a daughter nucleus during
$\alpha$-decay within the framework of our simple cluster model.

The fusion reactions between nuclei around the barrier are
strongly influenced by the coupling to both low-energy surface
vibration states and nucleon transfers
\cite{subfus-rev,CCFULL,denisov-tr,ccdef}. These two mechanisms of
sub-barrier fusion enhancement are considered in detail in the
construction of various models
\cite{subfus-rev,CCFULL,denisov-tr,ccdef}. Unfortunately, the
amplitude of this enhancement of sub-barrier fusion cross-section
varies depending on the model and various parameters. Moreover,
some of these parameters are often used for data fitting. However,
such coupling effects are small in the cases of stiff magic or
near-magic nuclei. The neutron transfer enhancement of sub-barrier
fusion cross section can be neglected when neutron transfer
channels with positive $Q$-value are absent \cite{denisov-tr}. We
chose the fusion reactions $\alpha$+$^{40}$Ca, $\alpha$+$^{59}$Co
and $\alpha$+$^{208}$Pb for evaluation of the $\alpha$-nucleus
potential, because the magic or near-magic nuclei $^4$He,
$^{40}$Ca, $^{59}$Co and $^{208}$Pb are very stiff and all 1- and
2-nucleon transfer channels have negative $Q$-values for these
reactions. Thus, in these reactions the values of sub-barrier
fusion cross-section evaluated by different models are very close
to each other, and we can make model-independent analysis of these
reactions without any fitting of additional parameters.
Fortunately, there are experimental data for fusion reactions
$\alpha$+$^{40}$Ca, $\alpha$+$^{59}$Co and $\alpha$+$^{208}$Pb
\cite{subfus-exp-ca1,subfus-exp-ca2,subfus-exp-co,subfus-exp-pb}.

Our cluster model for evaluation of $\alpha$-decay half-lives and
sub-barrier fusion reaction is presented in Sec. 2. The strategy
for $\alpha$-nucleus potential parameters searching is described
in Sec. 3. The discussion of the results and our conclusions are
given in Sec. 4.

\section{Model for $\alpha$-decay and sub-barrier fusion}

The $\alpha$-decay half-life $T_{1/2}$ is calculated as
\begin{eqnarray}
T_{1/2} = \hbar \ln(2)/\Gamma,
\end{eqnarray}
where $\Gamma$ is the total width of decay. The $\alpha$-particle
can be emitted from any point of the nuclear surface and to any
direction. It is obvious, however, that the $\alpha$-particle
emission in a direction normal to the nuclear surface is the most
profitable in terms of energy. Thus the total width is evaluated
by averaging partial widths (see also \cite{stewart,bm}).
Therefore the total $\alpha$-decay width is
\begin{eqnarray}
\Gamma = \frac{1}{4\pi} \int \gamma(\theta,\phi) d\Omega,
\end{eqnarray}
where $\gamma(\theta,\phi)$ is the partial width of
$\alpha$-emission in direction $\theta$ and $\phi$ and $\Omega$ is
the space angle. Note that similar averaging along the angle
$\Omega$ is also used for the evaluation of sub-barrier fusion
cross-sections between spherical and statically-deformed nuclei
\cite{subfus-rev,ccdef} (see below Eq. (11)).

The majority of the ground-state $\alpha$-emitters are spherical
nuclei or axial-symmetric nuclei with moderate quadrupole
deformation. Therefore we simplify the expression for total width.
It can be written as
\begin{eqnarray}
\Gamma= \int_0^{\pi/2} \gamma(\theta) \sin(\theta) d\theta  ,
\end{eqnarray}
where $\theta$ is the angle between the symmetry axis of
axially-symmetric deformed nuclei and the vector from the center
of the deformed nucleus to the emission point on the nuclear
surface. Due to the small or moderate values of the quadrupole
deformation of nuclei we neglect the difference between the
surface normal direction and $\theta$. It is obvious that $\Gamma
= \gamma(\theta) = \gamma(0)$ for spherical nuclei.

The width of $\alpha$-emission in direction $\theta$ is given by
the following:
\begin{eqnarray}
\gamma(\theta) = \hbar \; \xi \; t(Q,\theta,\ell),
\end{eqnarray}
where $\xi=\nu \cdot S$, $\nu$ is the frequency of assaults of a
$\alpha$-particle on the barrier, $S$ is the spectroscopic or
preformation factor, $t(Q,\theta,\ell)$ is the transmission
coefficient, which shows the probability of penetration through
the barrier, and $Q$ is the released energy at $\alpha$-decay.

The transmission coefficient can be obtained in the semiclassical
WKB approximation
\begin{eqnarray}
t(Q,\theta,\ell) = \{1
\;\;\;\;\;\;\;\;\;\;\;\;\;\;\;\;\;\;\;\;\;\;\;\;\;\;\;\;\;\;
\;\;\;\;\;\;\;\;\;\;\;\;\;\;\;\;\;\;\;\;\;\;\;\;\;\;\;\;\;\;
\\ + \left. \exp\left[\frac{2}{\hbar}
\int_{a(\theta)}^{b(\theta)} dr \sqrt{2\mu
\left(v(r,\theta,\ell,Q)-Q\right)} \right]\right\}^{-1}, \nonumber
\end{eqnarray}
where $a(\theta)$ and $b(\theta)$ are the inner and outer turning
points determined from the equations
$v(r,\theta,\ell,Q)|_{r=a(\theta),b(\theta)}=Q$, and $\mu$ is the
reduced mass. The $\alpha$-nucleus potential $v(r,\theta,\ell,Q)$
consists of Coulomb $v_C(r,\theta)$, nuclear $v_N(r,\theta,Q)$ and
centrifugal $v_\ell(r)$ parts, i.e.
\begin{eqnarray}
v(r,\theta,\ell,Q)=v_C(r,\theta) + v_N(r,\theta,Q) + v_\ell(r).
\end{eqnarray}

We propose that the parts of $\alpha$-nucleus potential be written
in the form
\begin{eqnarray}
v_C(r,\theta) =  \frac{2 Z e^2}{r} \left[1 + \frac{3R^2}{5r^2}
\beta Y_{20}(\theta) \right],
\end{eqnarray}
if $r \ge r_m$,
\begin{eqnarray}
v_C(r,\theta) &\approx& \frac{2 Z e^2}{r_m}
\left[\frac{3}{2}-\frac{r^2}{2r_m^2} \right. \\ & + & \left.
\frac{3R^2}{5r_m^2} \beta Y_{20}(\theta) \left(2-\frac{r^3}{r_m^3}
\right) \right], \nonumber
\end{eqnarray}
if $r\lesssim r_m$,
\begin{eqnarray}
v_N(r,\theta,Q) = V(A,Z,Q) / \{1
+\exp[(r-r_m(\theta))/d] \} , \\
v_\ell(r)= \hbar^2 \ell (\ell+1)/(2\mu r^2)  . \;
\end{eqnarray}
Here $A$, $Z$, $R$ and $\beta$ are, respectively, the number of
nucleons, the number of protons, the radius and the quadrupole
deformation parameter of the nucleus interacting with the
$\alpha$-particle, $e$ is the charge of proton, $Y_{20}(\theta)$
is the spherical harmonic function, and $V(A,Z,Q,\theta)$ and
$r_m(\theta)$ are, respectively, the strength and effective radius
of the nuclear part of $\alpha$-nucleus potential. The inner
turning point $a(\theta)$ is close to the touching point
$r_m(\theta)$, and therefore presentation of Coulomb field in the
form (8) at distances $r \lesssim r_m(\theta)$ ensures the
continuity of the Coulomb field and its derivative at $r=r_m$.

The trajectory of an $\alpha$-particle emitted from a deformed
nucleus is depicted by values of two coordinates $r$ and $\theta$.
An $\alpha$-particle emitted during the ground-state to the
ground-state transition has, as a rule, zero value of the orbital
momentum $\ell=0$ and negligible tangential velocity. Thus, we
disregard the small effects of variation of the angle $\theta$
during the barrier penetration in the case of $\alpha$-emission
from deformed nuclei. Therefore the action (5) related to the
sub-barrier penetrability depends only on $r$.

The sub-barrier fusion cross section between spherical projectile
and target nuclei with axial quadrupole deformation at collision
energy $E$ is equal to
\begin{eqnarray}
\sigma(E)=\frac{\pi \hbar^2}{2\mu E} \int_0^{\pi/2} \sum_\ell
(2\ell+1) t(E,\theta,\ell) \sin(\theta) d\theta
\end{eqnarray}
(see Refs. \cite{subfus-rev,CCFULL,ccdef,denisov-tr}). Here the
integration on angle $\theta$ is done for the same reason as for
Eqs. (2)-(3). The transmission coefficient $t(E,\theta,\ell)$ can
be obtained using various sub-barrier fusion models
\cite{subfus-rev,CCFULL,ccdef,denisov-tr}. We evaluated
$t(E,\theta=0,\ell)$ using the semiclassical WKB approximation (5)
in the case of collision between $\alpha$-particle and stiff magic
or near-magic spherical nuclei at collision energies $E$ below
barrier. The transmission coefficient is approximated by an
expression for a parabolic barrier \cite{subfus-rev,denisov-tr} at
collision energies higher then or equal to the barrier energy.

\section{Strategy of parameters searching}

We chose data for $T_{1/2}$ for 367 $\alpha$-decay transitions
between the ground states of parent and daughter nuclei from
tables in Refs. \cite{brazil,danevich,she-xu}. There are 166
even-even, 84 even-odd, 67 odd-even and 50 odd-odd parent nuclei
among these 367 $\alpha$-decay transitions.

The ground-state to the ground-state $\alpha$-transitions of
even-even nuclei took place at $\ell=0$. The value of $\ell$ can
be different from zero for the ground-state to the ground-state
transitions in odd or odd-odd nuclei. However we assume that all
$\alpha$-transitions between the ground states of parent and
daughter nuclei from Refs. \cite{brazil,danevich,she-xu} took
place at $\ell=0$, because information on value of $\ell$ is
absent in these data compilations. Similar approximation is also
used in Refs. \cite{mohr,blendowske,royer,brazil,she-xu}.

The $\alpha$-decay reaction $Q$-values were evaluated using recent
atomic mass data \cite{audi} or from \cite{she-xu} in the case of
superheavy nuclei.

The experimental data on static quadrupole deformation parameter
$\beta$ is taken from the RIPL-2 database \cite{ripl}. However, if
information on $\beta$ for a nucleus is not given in this
database, we picked up the value of $\beta$ from Ref. \cite{mnms}.

The data for sub-barrier fusion cross sections for reactions
$\alpha$+$^{40}$Ca, $\alpha$+$^{59}$Co and $\alpha$+$^{208}$Pb was
taken from
\cite{subfus-exp-ca1,subfus-exp-ca2,subfus-exp-co,subfus-exp-pb}.

We wanted to describe both the half-lives for 367 $\alpha$-decays
and fusion cross-section for reactions $\alpha$+$^{40}$Ca,
$\alpha$+$^{59}$Co and $\alpha$+$^{208}$Pb by using Eqs. (1)-(11).
By solving this task we determined the parameters $V(A,Z,Q)$,
$r_m(\theta)$ and $d$ of the nuclear part of the $\alpha$-nucleus
potential (9).

In determining the parameter values of the potential we took into
account the fact that various data are known with different
accuracy. The most accurate data is that of the ground-state to
the ground-state $\alpha$-transitions in even-even nuclei. The
data for similar transitions in odd or odd-odd nuclei are less
accurate as a rule, due to uncertainty of $\ell$, the level
schemes of parent and/or daughter nuclei and other reasons. The
data for fusion reactions is less accurate than data for
$\alpha$-decay half-lives as a rule. Furthermore, there are two
sets of data \cite{subfus-exp-ca1,subfus-exp-ca2} for fusion
reaction $\alpha$+$^{40}$Ca, which do not agree well with each
other.

Therefore we estimated the parameter values of the potential
starting from both data of the $\alpha$-decay half-lives $T_{1/2}$
of spherical and slightly deformed ($|\beta|\leq 0.05$) nuclei and
data of the fusion reaction cross-sections. Next we identified
specific features relating to the description of $T_{1/2}$ in
deformed nuclei.

At the very beginning of our study we tried to describe data for
$\log_{10}(T_{1/2})$ in even-even nuclei without taking into
account the fusion data. We tried to minimize the difference
\begin{eqnarray}
D_{e-e}=\sum_{\rm e-e \; nuclei}
[\log_{10}(T_{1/2}^{theor.})-\log_{10}(T_{1/2}^{expt.})]^2,
\end{eqnarray}
where $T_{1/2}^{theor.}$ and $T_{1/2}^{expt.}$ are theoretical and
experimental values of half-lives respectively. However we could
not fix the parameters of $\alpha$-nucleus potential, because it
is possible to describe $\log_{10}(T_{1/2})$ in even-even nuclei
with similar values of root mean square error using very different
values of nuclear potential strength $V(A,Z,q,\theta)$,
diffuseness $d$, radii $r_m$ and $\xi$. Note that this situation
is typical. The values of $\log_{10}(T_{1/2})$ are well described
in Refs. \cite{back,blendowske,mohr,royer,brazil,poenaru};
however, the values of potential strength, diffuseness, radii and
$\xi$ in these Refs. vary by large intervals.

Subsequently, we tried to describe simultaneously data for
$\log_{10}(T_{1/2})$ in even-even, odd and odd-odd nuclei as well
as for the sub-barrier fusion data for the reaction
$\alpha$+$^{208}$Pb. We paid special attention to the description
of $\log_{10}(T_{1/2})$ in even-even data, and therefore for our
parameter searching we tried to minimize the function
\begin{eqnarray}
100 D_{e-e}+D_{e-o}+D_{o-e}+D_{o-o}\\ +\sum_{k}
[\sigma^{theor.}_{fus}(E_k)-\sigma^{expt.}_{fus}(E_k) ]^2
\nonumber
\end{eqnarray}
Here $D_{e-o}$, $D_{o-e}$, $D_{o-o}$ are the differences similar
to (12) for even-odd, odd-even and odd-odd data sets
correspondingly, $\sigma^{theor.}_{fus}(E_k)$ and
$\sigma^{expt.}_{fus}(E_k)$ are theoretical and experimental
values of fusion cross section at energy $E_k$ for reaction
$\alpha$+$^{208}$Pb respectively. We took into account all data on
cross-sections of the reaction $\alpha$+$^{208}$Pb. The values of
fusion cross-section are expressed in millibarns during the
minimization procedure. The experimental errors in subbarrier
fusion cross-section values are raised with reduction of collision
energy especially in the sub-barrier region.
\cite{subfus-exp-ca1,subfus-exp-ca2,subfus-exp-co,subfus-exp-pb}.
This is taken into account in Eq. (13).

The quality of description of $\alpha$-decay half-lives is weakly
influenced by the last term in Eq. (13). However, by using this
approach we remove the freedom in the choice of parameter $\xi$.
The value of parameter $\xi$ is coupled to the parameters of the
potential in this step. Nevertheless, we cannot strictly determine
the parameters of the potential, because we may describe target
data sets with comparable values of root mean square error using
very different values of nuclear potential strength
$V(A,Z,q,\theta)$.

The additional demand that the parameters describe the
cross-section data for fusion reactions $\alpha$+$^{40}$Ca and
$\alpha$+$^{59}$Co as well as the previous target data sets, make
it possible to fix the parameters of the nuclear part of the
potential between $\alpha$-particle and spherical nuclei. Note
that the data sets for fusion cross-sections for reactions
$\alpha$+$^{40}$Ca and $\alpha$+$^{59}$Co are known to have lower
accuracy than other data sets, and therefore for parameter
searching we give very small factors to the terms related to the
difference between theoretical and experimental cross-sections
which we add to the function (13). These small factors reasonably
diminish the influence of fusion data sets for the reactions
$\alpha$+$^{40}$Ca and $\alpha$+$^{59}$Co.

The obtained values of parameters are
\begin{eqnarray}
V(A,Z,Q)=-[30.275  - 0.45838 Z/A^{1/3} \\
+ 58.270 I  - 0.24244 Q ], \nonumber \\
r_m =1.5268 + R, \\
R=R_p (1+3.0909/R_p^2) + 0.12430 t,\\
R_p=1.24 A^{1/3} (1 + 1.646 /A - 0.191 I), \\
t=I-0.4 A/(A+200),  \\
d=0.49290 ,\\
\xi = (6.1814+0.2988 A^{-1/6}) 10^{19} {\rm s}^{-1},
\end{eqnarray}
where $I=(A-2Z)/A$. Here we use a method for determining the
radius parameters similar to that used in Ref. \cite{denisov-pot}
for evaluation of nuclear part of the potential between two heavy
spherical nuclei.

The quality of description of fusion reactions $\alpha$+$^{40}$Ca
and $\alpha$+$^{59}$Co is degraded for other sets of potential
parameters.

The density distribution of a deformed nucleus is described by
deformation parameter $\beta$ and angle $\theta$. Therefore, the
potential between $\alpha$-particle and deformed nucleus should
depend on deformation $\beta$ and angle $\theta$, because the
nuclear part of the potential is strongly linked to nucleon
density in the double-folding model \cite{denisov-pot,dn}. It is
natural that the parameter values determining $\alpha$-nucleus
potential do not change with transition from a spherical to a
deformed nucleus. Therefore, the angular dependence of the
potential between an $\alpha$-particle and a deformed nucleus can
be linked to the density distribution of a deformed nucleus. As a
result, the relation between deformation of the nucleus and
angular dependence of the nuclear part of $\alpha$-nucleus
potential can be associated with the radius parameters, i.e.
\begin{eqnarray}
r_m(\theta) =1.5268 + R(\theta) , \\
R(\theta)=R(1+\beta Y_{20}(\theta)).
\end{eqnarray}

The height of the barrier between the $\alpha$-particle and the
deformed nucleus and the inner turning point $a(\theta)$ become
strongly dependent on $\theta$. The dependence of potential on
angle $\theta$ is similar to that in the case of heavier nuclei
\cite{dn}. The barrier between an $\alpha$-particle and a prolate
nucleus is lower and thinner at $\theta=0^\circ$ than at
$\theta=90^\circ$. Due to this the transmission coefficient
$t(q,\theta=0^\circ,\ell)$ in a prolate nucleus is larger than the
transmission coefficient in a spherical nucleus where other
parameters have the same values. As a the result the evaluated
half-lives of prolate nuclei are strongly reduced when we take
into account the deformation of nuclei (see also Fig. 1 in Ref.
\cite{delion1}). However, the $\alpha$-decay hindrance caused by
deformation of nuclei \cite{bm} strongly affects $\alpha$-decay
half-lives.

For the sake of fitting data for half-lives of deformed nuclei we
introduce a deformation dependence parameter $\xi$
\begin{eqnarray}
\xi = (6.1814+0.2988 A^{-1/6}) 10^{19}  \exp{(-13.116 \beta)} \;
{\rm s}^{-1}. \;
\end{eqnarray}

Note that $\xi$ is the product of the assault frequency and the
formation probability of $\alpha$-particle cluster in the parent
nucleus. The exponential factor in Eq. (23) reflects the fact that
deformation strongly influence the formation probability of
$\alpha$ cluster in the parent nucleus. Note that the deformation
or angular dependency of $\alpha$-nucleus potential near inner
turning point $a(\theta)$ may slightly affect the assault
frequency due to the variation of $\alpha$-nucleus potential. This
exponential factor reflects the hindrance of the $\alpha$-cluster
formation in prolate nuclei and the enhancement of the one in
oblate nuclei. It is significant to note that the exponential
factor depended on deformation also is adopted in detailed
microscopical $\alpha$-decay theories
\cite{stewart,delion1,delion2,bm}.

It is obvious that violation of spherical symmetry leads to
modification of $\alpha$-particle localization on the surface of
the deformed nucleus. Due to static quadrupole deformation the
strong coupled-channel effect between outgoing waves with
$\ell=0,2,4$ can also attenuate $\ell=0$ $\alpha$-particle
transitions in prolate nuclei \cite{stewart,delion1,delion2}. Note
that the transmission coefficient and half-life are strongly
reduced with rising of $\ell$ due to the centrifugal term (6),
(10). For these reasons the introduction of the exponential factor
in Eq. (23) is natural. However it is desirable to discuss in
detail this exponential factor in the light of microscopical
considerations.

Note that $t(q,\theta=0^\circ,\ell)> t(q,\theta=90^\circ,\ell)$ in
prolate nuclei. Therefore if we propose the independence of
$\alpha$-cluster formation probability on the angle $\theta$, then
the $\alpha$-particles should be emitted mainly at angle
$\theta=0^\circ$ (see also
\cite{stewart,delion2,aleshin,alexander,huizenga2,severijns}). A
similar effect is well-known in the case of sub-barrier fusion
reactions between spherical and deformed nuclei
\cite{subfus-rev,ccdef}.

\section{Results and discussions}

\begin{figure*}
\hspace{-8.5mm}
\includegraphics[width=16.2cm]{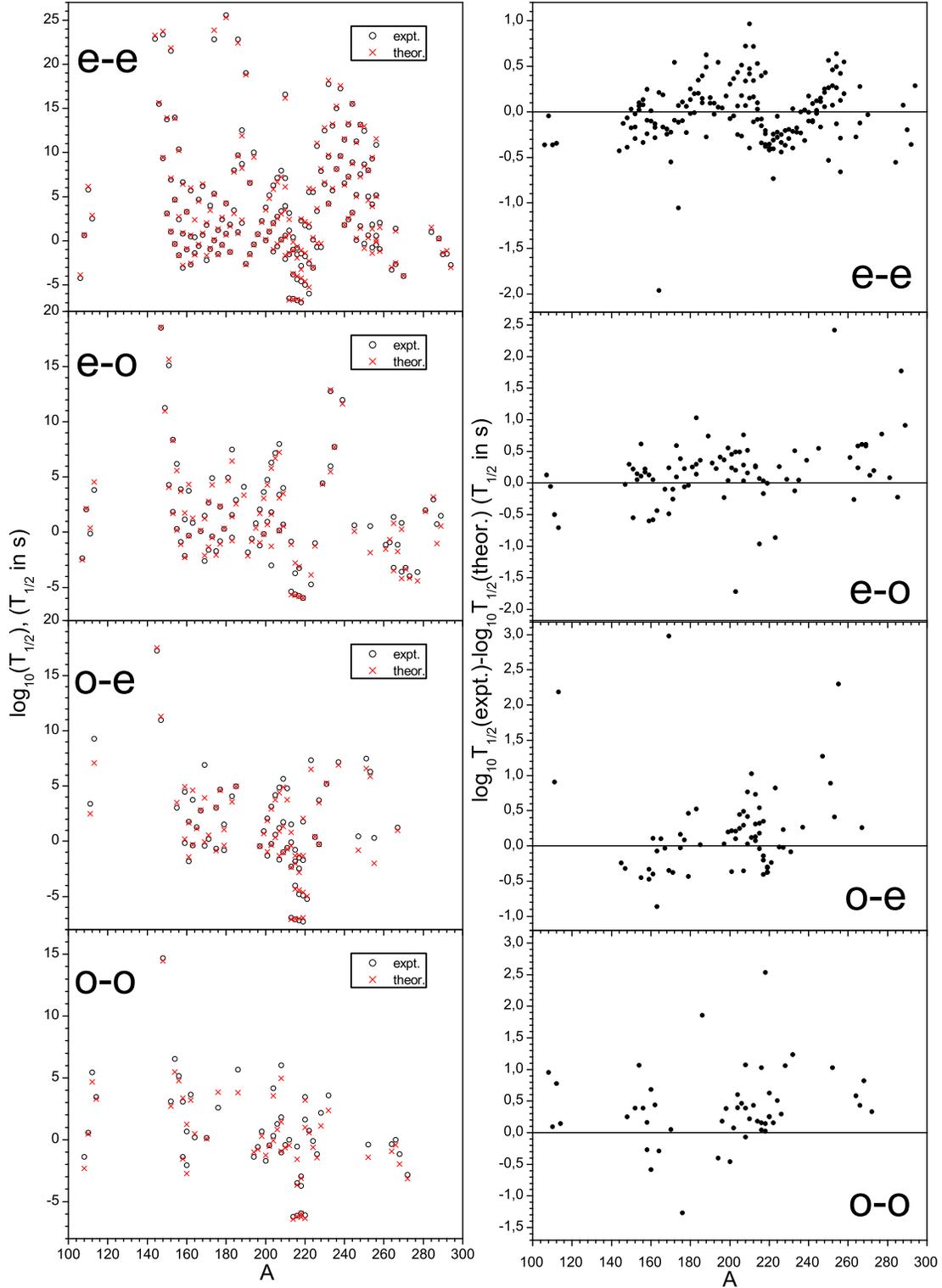}
\caption{Left panels: The experimental (circles)
\cite{brazil,danevich,she-xu} and theoretical (crosses) values of
$\log_{10}(T_{1/2})$ for $\alpha$-decays in even-even (e-e),
even-odd (e-o), odd-even (o-e) and odd-odd (o-o) parent nuclei.\\
Right panels: Dots represent the difference between the
experimental and theoretical values of $\log_{10}(T_{1/2})$ for
$\alpha$-decays in even-even (e-e), even-odd (e-o), odd-even (o-e)
and odd-odd (o-o) parent nuclei.}
\end{figure*}

\begin{figure*}
\includegraphics[width=17.0cm]{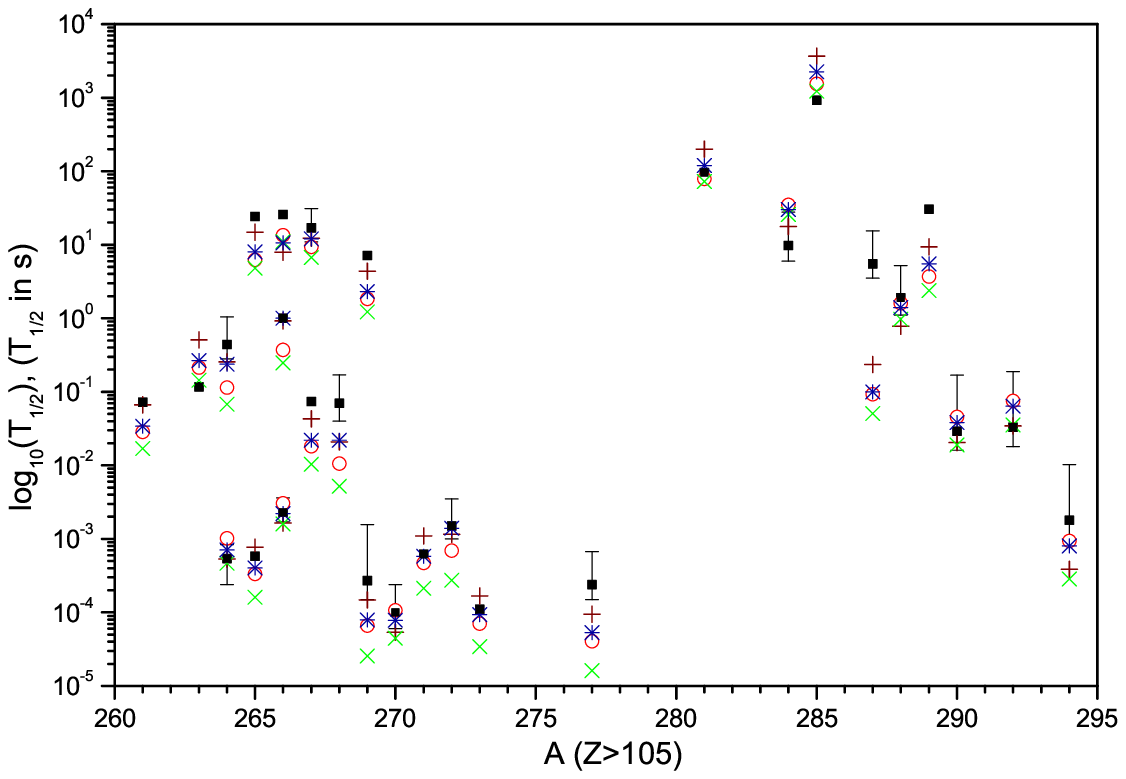}
\caption{The experimental and theoretical values of
$\log_{10}(T_{1/2})$ for superheavy region. Squares  with error
bars are data from \cite{brazil,she-xu}, circles are theoretical
values obtained by using Eqs. (1)-(9) and (11)-(17), plus and
cross signs are the values obtained by using empirical relations
from Refs. \cite{royer} and \cite{sss} respectively, and stars are
the results of calculations from Ref. \cite{she-xu}. }
\end{figure*}

The results of our evaluations of various data are presented in
Figs. 1-3. We start our discussion with detailed consideration of
the $\alpha$-decay half-lives in nuclei.

\subsection{$\alpha$-decay half lives}

The $\alpha$-decay half-lives evaluated by using Eqs. (1)-(10),
(14)-(23) agree well with experimental data (see Fig. 1-2). The
values of half-lives are scattered over an extremely wide range
from $\approx 10^{-7}$s to $\approx 10^{+25}$s. The $\alpha$-decay
half-lives are very nicely described in the case of even-even
parent nuclei. We see in Fig. 1 that the difference between
theoretical and experimental values of $\log_{10}T_{1/2}$ are
smaller than 0.5 for most of cases of even-even, even-odd and
odd-even nuclei. The evaluation of odd-odd nuclei is a little bit
worse.

The $\alpha$-particles emitted from superheavy elements are
considered in recent Refs. \cite{royer,she,she-xu,sss}. In Fig. 2
we present the results for $\log_{10}(T_{1/2})$ of superheavies
using our model and other approaches \cite{royer,she-xu,sss}. Our
results and those from Ref. \cite{she-xu} are obtained by
different cluster model approaches to the $\alpha$-decay, while
results from Refs. \cite{royer,sss} are evaluated with the help of
various empirical relations. The empirical relations used in Refs.
\cite{poenaru,royer,sss,brown} couple $\log_{10}(T_{1/2})$,
$\alpha$-particle $Q$-value, mass and charge of parent nuclei by
simple functional expressions. As a rule, empirical relations are
derived by using a pure Coulomb picture of $\alpha$-decay, which
neglects both the nuclear force between $\alpha$-particle and
daughter nucleus and the deformation of daughter nucleus
\cite{brown}. Nevertheless the empirical relations are often used
to estimate of $\log_{10}(T_{1/2})$ due to their simplicity. The
empirical relation from Ref. \cite{sss} was derived especially for
description of $\log_{10}(T_{1/2})$ in heavy and superheavy
nuclei. In Ref. \cite{royer} four empirical relations for
even-even, even-odd, odd-even and odd-odd $\alpha$-decaying nuclei
are established.

We see in Fig. 2 that our approximation describes
$\log_{10}(T_{1/2})$ for the superheavy region  better than the
empirical relation from Ref. \cite{sss} and worse than the set of
empirical relations from Ref. \cite{royer}. The cluster theory
proposed in Ref. \cite{she-xu} describes well $\log_{10}(T_{1/2})$
in this region too. However, a renormalization factor for the
nuclear part of $\alpha$-nucleus potential is used for each decay
case in Ref. \cite{she-xu}. In contrast to this, our model
describes well the half-lives of nuclei in a very wide region with
the same set of potential parameters (see Fig. 1 and 2), and takes
into account the deformation effects which are omitted in other
approaches.

\begin{figure}
\hspace{-.6cm}\includegraphics[width=9.2cm]{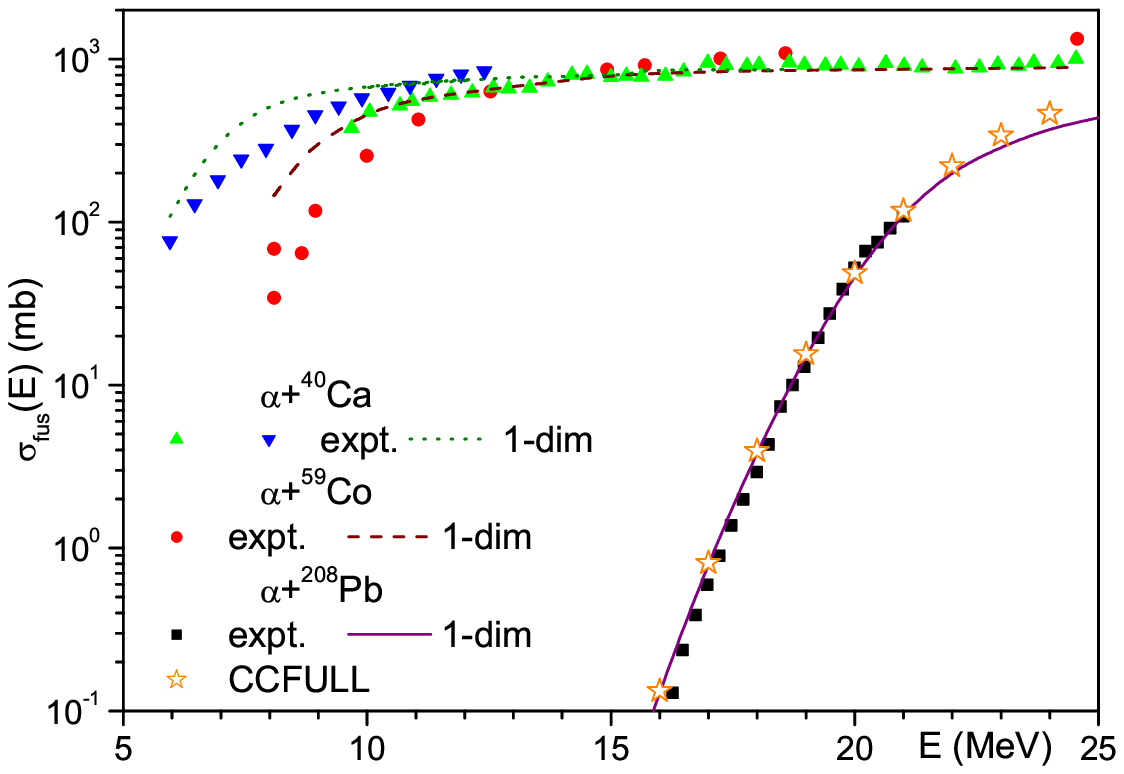}
\caption{The experimental and theoretical values of fusion
cross-section for reactions $\alpha$+$^{40}$Ca, $\alpha$+$^{59}$Co
and $\alpha$+$^{208}$Pb. Squares are data for reaction
$\alpha$+$^{208}$Pb from Ref. \cite{subfus-exp-pb}, circles are
data for reaction $\alpha$+$^{59}$Co from Ref.
\cite{subfus-exp-co}, up- and down-pointing triangles are data for
reaction $\alpha$+$^{40}$Ca from Refs. \cite{subfus-exp-ca2} and
\cite{subfus-exp-ca1}, respectively. Lines are results of
calculations obtained by using Eqs. (5)-(11), (14)-(19), and stars
are result of calculations using CCFULL code \cite{CCFULL}.}
\end{figure}

\subsection{Fusion cross-sections}

The fusion cross-sections evaluated using Eqs. (5)-(11), (14)-(19)
for reactions $\alpha$+$^{40}$Ca, $\alpha$+$^{59}$Co and
$\alpha$+$^{208}$Pb are compared with experimental data
\cite{subfus-exp-ca1,subfus-exp-ca2,subfus-exp-co,subfus-exp-pb}
in Fig. 3. We see that our model very accurately describes the
data for fusion reaction $\alpha$+$^{208}$Pb. The data for
reactions $\alpha$+$^{40}$Ca and $\alpha$+$^{59}$Co are also well
described by our model.

Our model for evaluation of the fusion cross-section between an
$\alpha$-particle and a spherical nucleus is one-dimensional. As
we pointed in the Introduction, the coupled-channel effects are
very important for the nuclear fusion reaction around the barrier
\cite{subfus-rev,CCFULL,denisov-tr,ccdef}. Thus, we also made
result of the coupled-channel calculation of the fusion
cross-section for reaction $\alpha$+$^{208}$Pb by using CCFULL
code \cite{CCFULL}, and the results are presented in Fig. 3. The
effects of nonlinear coupling of the low-energy surface
vibrational states in all orders are taken into account in this
code. The CCFULL calculation uses the same $\alpha$-nucleus
potential as in the case of one-dimensional calculation. The
values of excitation energies and dynamic surface deformations are
taken from \cite{ripl}. As we can see in Fig. 3, the agreement
between our one-dimensional and coupled-channel calculations is
rather good. The good agreement between CCFULL and one-dimensional
calculations ca be attributed to both the high stiffness of
double-magic nuclei participating in this reaction and the smaller
values of $\alpha$-nucleus potential and its derivative than in
the case of a more symmetric colliding system.

\subsection{Comparison with other approaches}

The value of the depth $V(A,Z,Q)$ (14) of the nuclear part of the
$\alpha$-nucleus potential evaluated in our model is smaller than
that typically obtained using data for high energy reactions or
some calculation from M3Y nucleon-nucleon forces
\cite{pot-nolte,mohr,pot-avrigeanu}. However, the depth of the
potential is unimportant in the analyzing of reaction data around
the barrier. For example, it is possible to obtain
$\alpha$-particle elastic scattering and total reaction cross
section data from either the deep or shallow nuclear part of the
$\alpha$-nucleus potential using the optical model
\cite{huizenga}.

Small values of the depth of the nuclear part of the
$\alpha$-nucleus potential are derived in the analysis of the
low-energy data as a rule. Thus, good estimation of the
$\alpha$-decay half-lives for superheavy nuclei in Ref.
\cite{she-xu} (see also Fig. 2) is obtained by strong reduction of
the nuclear part of the potential calculated from the M3Y
nucleon-nucleon force. The strength of the nuclear part of
$\alpha$-nucleus potential obtained in Ref. \cite{royer} is even
smaller than that obtained by our approach (see Fig. 4).

The $\alpha$-decay and subbarrier fusion are slow processes. The
strong repulsion arises between nuclei at low collision energies
due to the Pauli principle \cite{wt,khoa}. As a result, the
$\alpha$-nucleus potential evaluated using the M3Y nucleon-nucleon
force becomes shallower due to the Pauli repulsion. The depth of
Woods-Saxon type potential is approximately less by half than the
depth of M3Y type potential evaluated for the same
$\alpha$-nucleus system, when the both potentials are close to
each other at large distances \cite{khoa}. Therefore the shallow
$\alpha$-nucleus potential is reasonable for study of both
$\alpha$-decay and subbarrier fusion.

Our value of potential diffuseness (19) is smaller then the ones
from Refs.
\cite{huizenga,pot-nolte,pot-avrigeanu,blendowske,royer}. The
diffuseness of potential in the double-folding model is related to
both the diffuseness of density distribution of interacting nuclei
and the diffuseness of nucleon-nucleon force. The diffuseness of
$\alpha$-particles is small because density distribution of
$\alpha$-particles is close to Gaussian (see examples in
\cite{pot-demetriou,pot-avrigeanu}). The diffuseness of density
distribution in heavy nuclei does not depend on mass number as a
rule. Therefore, the mass independent small value of diffuseness
obtained in our study is reasonable.

The expression for effective nuclear radius $R$ is close to the
one used in \cite{denisov-pot} for evaluation of the potential
between two nuclei. The expression for proton radius $R_p$ (14) is
proposed in \cite{pomorska}. The value of $\alpha$-particle
radius, $1.5268$ fm (see Eq. (15)), is very close to the
experimental value 1.57$\pm$0.05 fm \cite{tanihata}. Note that
during potential parameter searching, the value of
$\alpha$-particle radius is scanned in the interval from 0.5 fm to
2.0 fm.

The comparison of the nuclear part of the potentials at distance
larger than touching point evaluated using our and other
\cite{pot-nolte,pot-avrigeanu,blendowske,royer} approaches is
presented in the Fig. 4. The potentials for $\alpha+^{40}$Ca and
$\alpha+^{208}$Pb are evaluated at $E=20$ MeV. We see in this
figure that our potential is very close to the potentials from
Refs. \cite{pot-nolte,pot-avrigeanu,blendowske} near the touching
points for $\alpha+^{208}$Pb. At larger distances between nuclei
our potential is less attractive then others. However, our
potential is more attractive than any other
\cite{pot-nolte,pot-avrigeanu,blendowske,royer} for
$\alpha+^{40}$Ca case, see Fig. 4. Here we should note that
potential parameterizations from Refs.
\cite{pot-nolte,pot-avrigeanu,blendowske} are obtained by using
data of interactions between $\alpha$ and medium or heavy nuclei.
The potential evaluated in Ref. \cite{royer} is less attractive
then others at distances close to the touching point.

\subsection{Fusion for superheavy element spectroscopy studies}

\begin{figure}
\hspace{-.6cm}\includegraphics[width=9.2cm]{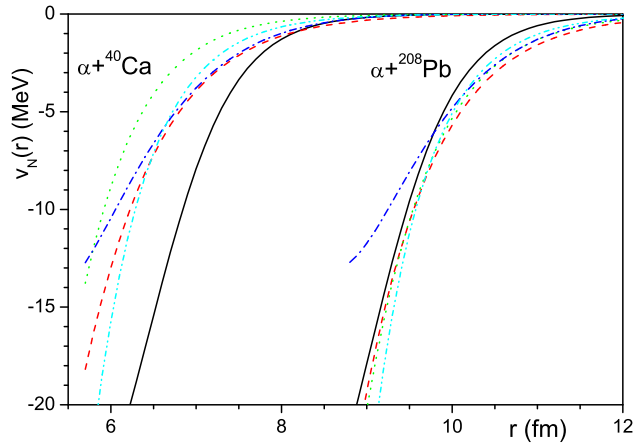}
\caption{The nuclear part of potentials between $\alpha$-particle
and $^{40}$Ca or $^{208}$Pb. The solid, dash, dots, dash-dot and
dash-dot-dot lines are evaluation results using our model and
parametrizations from Refs. \cite{pot-nolte},
\cite{pot-avrigeanu}, \cite{royer} and \cite{blendowske},
respectively.}
\end{figure}

Recently the fusion reactions between heavy nuclei have been used
for spectroscopy studies of superheavy elements
\cite{she,herzberg}. The cross-sections of reactions used for this
purpose are very small \cite{herzberg}, because of
compound-nucleus formation hindrance \cite{armbruster}. The fusion
hindrance between $\alpha$-particle and very heavy nucleus on the
other hand is absent, because of the small value of $2 \cdot Z$,
where 2 is the charge of $\alpha$-particle and $Z$ is the charge
of heavy nucleus. Therefore fusion reactions between
$\alpha$-particle projectiles and very heavy target nuclei can be
also used for heavy nucleus spectroscopy studies.

It is possible to make spectroscopy studies of a compound nucleus
on modern facilities if the compound-nucleus cross-section is
larger than 0.2 $\mu$b \cite{herzberg}.

We present the fusion cross-section for the reaction
$\alpha$+$^{252}$Cf=$^{256}$Fm evaluated by using Eqs. (5)-(11),
(14)-(19) in Fig. 5. The compound-nucleus cross-section shown in
Fig. 5 is evaluated for different proposed shapes of $^{252}$Cf.
The spherical $^{252}$Cf cross-section is smaller than the
deformed one. Similar results are also found for the fusion
reactions between heavier projectiles and lighter targets
\cite{subfus-rev,ccdef}. The value of quadrupole deformation
parameter for $^{252}$Cf is taken from \cite{ripl}. From Fig. 5 we
see that it is possible to make spectroscopy studies of $^{256}$Fm
at collision energies $E \gtrsim 16$ MeV. The excitation energies
of $^{256}$Fm compound nucleus at such collision energies are
moderate.

Note that we evaluate the fusion cross-section by using the
one-dimensional WKB approximation for reaction between spherical
and deformed nuclei. Coupling to the low-energy vibrational states
in $^{252}$Cf can slightly increase the cross-section values
presented in Fig. 5. Due to this, our estimation of cross-section
is the lower limit of the cross-section.

\begin{figure}
\hspace{-.6cm}\includegraphics[width=9.2cm]{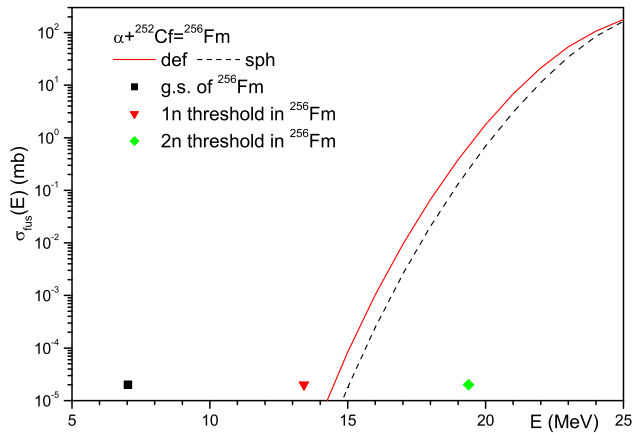}
\caption{The fusion cross-section for
$\alpha$+$^{252}$Cf=$^{256}$Fm reactions. The solid and dash lines
are fusion cross-sections in the cases of deformed and spherical
shape of $^{252}$Cf respectively. The square marks the collision
energy at which $^{256}$Fm is formed in the ground state. The
triangle and diamond are threshold energies for 1 and 2 neutron
emission from $^{256}$Fm, respectively.}
\end{figure}

The spectroscopy studies of $^{255}$Fm can be made after $1n$
emission from $^{256}$Fm. The neutron separation energy in
$^{256}$Fm is $E_{1n}=6.4$ MeV \cite{audi} and the experimental
fission barrier is $E_f=4.8$ MeV \cite{msi}. The compound-nucleus
survival probability $W$ is related with neutron emission
$\Gamma_n(E^\star)$ and fission $\Gamma_f(E^\star)$  widths, i.e.
$W\approx\Gamma_n(E^\star)/\Gamma_f(E^\star)$ \cite{dh}. These
widths are proportional to the level densities ratio, and
therefore
\begin{eqnarray}
W(E^\star) \propto \nonumber \;\;\;\;\;\;\;\;\;\;\;\;\;\;\;\;\;\;\;\;\;\;\;\;\;\;\;\;\;\;\;\;\;\;\;\;\;\;\;\;\;\;\;\;\;\;\;\;\;\;\;\;\;\;\;\;\;\;\;\; \\
\exp{[2(a(E^\star-E_n))^{1/2}-2(a(E^\star-E_f))^{1/2}]},
\end{eqnarray}
where $E^\star$ is the excitation energy of the compound nucleus
and $ a=0.114 A+0.162 A^{2/3}$ is asymptotic level density
parameter in a nucleus with $A$ nucleons \cite{ignatyuk}. We see
in this figure that the cross section for $^{256}$Fm formation at
an energy level just below the $2n$ emission threshold ($E\approx
19$ MeV) is close to 0.4 mb. The compound-nucleus survival
probability $W$ at this excitation energy is $W \approx 0.02$.
Therefore the cross-section of the $\alpha$+$^{252}$Cf=$^{255}$Fm
$+1n$ reaction is $\sigma_{fus}(E) W(E-Q)\approx 8 \mu$b. Here we
take into account that the excitation energy of compound-nucleus
$^{256}$Fm formed in the fusion reaction $\alpha$+$^{252}$Cf at
collision energy $E$ is $E^\star = E-Q$. The value of the cross
section is high enough for spectroscopic studies of $^{255}$Fm.

The fusion reactions induced by $\alpha$-particles have $1n$
channels as a rule. In contrast to this, more neutrons are
generally evaporated in the reactions between heavier projectiles
and lighter targets leading to the same compound-nucleus formed in
reactions induced by an $\alpha$-projectile. However, the poor
availability of heavy targets limits the use of $\alpha$-capture
reactions. Nevertheless, the cross sections of these reactions are
relatively high, and therefore these reactions are attractive for
spectroscopic studies.

In conclusion, we determined the $\alpha$-nucleus potential by
using the data for $\alpha$-decay half-lives and sub-barrier
fusion reactions. The data for $\alpha$-decay half-lives play
principal role at potential evaluation. The data for
$\alpha$-decay half-lives in spherical and deformed nuclei and for
sub-barrier fusion reactions $\alpha$+$^{40}$Ca,
$\alpha$+$^{59}$Co, $\alpha$+$^{208}$Pb are well described by our
model. We showed that it is possible to use $\alpha$-nucleus
fusion reactions for the spectroscopic studies of very heavy
nuclei. The sub-barrier $\alpha$-capture reactions can be
fruitfully used for the spectroscopy studies of very heavy nuclei.

V.Yu. Denisov would like to thank the Japan Society for the
Promotion of Science for its financial support.

\end{document}